# Low-threshold stimulated emission using colloidal quantum wells

Chunxing She[1#], Igor Fedin[1#], Dmitriy S. Dolzhnikov[1], Arnaud Demortière[2,3], Richard D. Schaller[4,5], Matthew Pelton[4,6]*, and Dmitri V. Talapin[1,4]*

[1] Department of Chemistry and James Frank Institute, University of Chicago, Illinois 60637, USA

[2] Materials Science Division, Argonne National Laboratory, Argonne, Illinois 60439, USA

[3] Physics Department, Illinois Institute of Technology, Chicago, IL 60616, USA

[4] Center for Nanoscale Materials, Argonne National Laboratory, Argonne, Illinois 60439, USA

[5] Department of Chemistry, Northwestern University, Evanston, IL 60208, USA

[6] Department of Physics, University of Maryland, Baltimore County, Baltimore, MD 21250, USA

[#] These authors contribute to this work.

[*] e-mail: mpelton@umbc.edu; dvtalapin@uchicago.edu

**Semiconductor nanocrystals can be synthesized using inexpensive, scalable, solution-based techniques[1], and their utility as tunable light emitters has been demonstrated in various applications, including biolabeling[2] and light-emitting devices[3]. By contrast, the use of colloidal nanocrystals for optical amplification[4] and lasing[5] has been limited by the high input power densities that have been required. In this work, we show that colloidal nanoplatelets (NPLs) produce amplified spontaneous emission (ASE) with pump-fluence thresholds as low 6 µJ/cm$^2$ and gain as high as 600 cm$^{-1}$, both a 4-fold improvement over the best reported values for colloidal nanocrystals; in addition, gain saturation occurs at pump fluences two orders of magnitude higher than the ASE threshold. We attribute this exceptional performance to large optical cross-sections, slow Auger recombination rates, and the narrow emission linewidth of the NPL ensemble. The NPLs bring the advantages of quantum wells as an optical gain medium to a colloidal system, opening up the possibility of producing high-efficiency, solution-processed lasers.**

Ever since highly luminescent colloidal semiconductor nanocrystals, or quantum dots (QDs), were first synthesized[6], there has been an abiding interest in using them laser gain media, due to their tunable emission wavelengths, low cost, and solution processability. However, nearly a decade passed before the first demonstrations of optical gain and lasing[4,5,7,8] from colloidal QDs. These were achieved only under pulsed excitation at high energy densities: thresholds for amplified spontaneous emission (ASE) in films of QDs were on the order of 1 mJ/cm$^2$.

The high thresholds were attributed to Auger processes: optical gain requires excitation of more than one exciton in each QD, but multiple excitons undergo rapid, non-radiative Auger recombination[9]. Attempts to reduce thresholds therefore focused on reducing the effects of Auger recombination. For example, charged QDs[10] or carefully engineered core-shell QDs[11] can have spectrally separated absorption and emission energies, enabling gain without excitation of multiple excitons. Alternatively, nanocrystal heterostructures can be engineered for reduced Auger recombination rates[12]. Reduced Auger rates and lower thresholds were also obtained for QDs embedded in thick shells with different shapes, including rods[13], tetrapods[14], and spheres (or "giant" QDs)[15]. In this case, optical absorption at the excitation energy is also increased[15], leading to thresholds as low as ~26 µJ/cm$^2$. The lower threshold, however, comes at the expense of the maximum obtainable gain, because the thick shells reduce the maximum exciton density that can be obtained in a close-packed film of these nanocrystals.

In contrast, low-threshold lasing has long been obtained using epitaxially grown quantum wells (QWs)[16]. In these structures, carriers are confined in only one dimension, and Auger recombination rates at ASE thresholds are expected to be lower than in QDs. However, the techniques generally used to produce QWs, such as molecular beam epitaxy[17], are expensive and low-throughput, and only a relatively small number of QWs can be stacked above one another, limiting the volume of gain medium that can be produced. Recently, methods have been developed for the colloidal synthesis of thin, flat semiconductor nanocrystals, including nanoribbons[18,19] and nanoplatelets (NPLs)[20,21]. Carriers in these flat nanocrystals are confined in only one dimension, making them the colloidal equivalent of QWs. Time-resolved optical measurements have shown that high-energy carriers in NPLs relax to the band edge on time scales much faster than recombination times[22], as needed for stimulated emission and lasing.

In this work, we demonstrate ASE using close-packed films of NPLs, and show that ASE thresholds are lower and gain saturation is higher than for other semiconductor nanocrystals. We study both CdSe NPLs and CdS/CdSe/CdS shell/core/shell NPL heterostructures, illustrated in Figure 1a. We start with CdSe NPLs with lateral dimensions of 27 ± 3 nm and 6.8 ± 1.5 nm (see Figure 1c). The thickness of the platelets is 1.2 nm, equivalent to 4 complete monolayers (MLs) of CdSe with an additional layer of Cd atoms, so that both sides of the NPL are Cd-terminated (see Figure 2). The NPLs form stable colloidal solutions (see inset in Figure 1c), which show very narrow excitonic peaks in absorption (see Figure 1b) and emission (see the Supporting Information), with an emission maximum at 2.42 eV (512 nm)[20,21].

NPL heterostructures, shown in Figure 1d, are produced by growing CdS shells with thickness from 1 to 8 MLs on either side of the CdSe NPLs using the technique of colloidal atomic layer deposition (c-ALD)[23]. We refer to the resulting structures as $x$CdS/CdSe/$x$CdS, where $x$ corresponds to the thickness of the shell. (For the core, there are four Se layers and five Cd layers.) As shown in Figure 1b, the absorption spectra of the NPLs shift continually to lower energy as $x$ increases, due to the extension of the electron wavefunction into the CdS shell (see Supplementary Figure S1 for more examples). The emission spectra show the same shifts (see

Supplementary Figure S2). The c-ALD process enables the growth of an atomically flat CdS shell with ML precision over thickness[23]. This is illustrated in Figure 2 for a single 8CdS/CdSe/8CdS NPL. (An NPL with a thick shell was chosen in this case to clearly illustrate the shell/core/shell structure.)

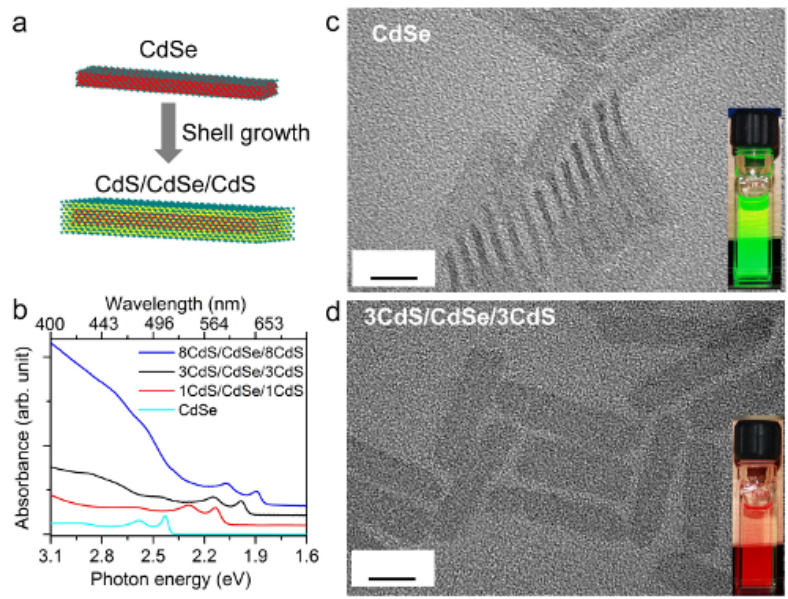

**Figure 1. CdSe and CdS/CdSe/CdS nanoplatelets (NPLs)** (a) Schematic structures of NPLs. Cd, Se, and S atoms are in blue, red, and green, respectively. (b) Absorption spectra of CdSe and $x$CdS/CdSe/$x$CdS ($x$ = 1, 3 and 8) NPLs in hexane. (c-d) Transmission electron microscope images of (c) CdSe and (d) 3CdS/CdSe/3CdS NPLs, viewed from the top. Scale bars are 10 nm. Insets are photographs of luminescence from solutions of NPLs in hexane.

In order to observe ASE, we deposited CdSe NPLs on glass substrates by spin coating; the resulting films are optically uniform and transparent, with a thickness of 168 ± 19 nm, a refractive index at 632 nm of 1.7 ± 0.2, and thus a packing density of 50±15%. We excited the films with frequency-doubled pump pulses from an amplified Ti:Sapphire laser system, focused to a stripe along the sample. When the pump fluence exceeds a certain threshold, a narrow peak due to ASE appears on the lower-energy side of the broader photoluminescence band, and the intensity of this peak increases rapidly with pump fluence. This is shown in Figure 3a for a film of CdSe NPLs that has a threshold fluence of 39 $\mu J/cm^2$. This ASE threshold is comparable to the lowest reported values for complex QDs specially optimized for low-threshold lasing[12,15].

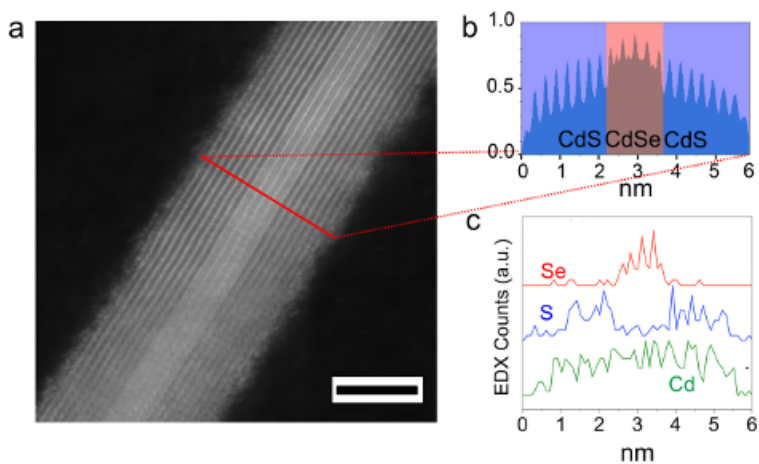

**Figure 2. Nanoplatelet structure.** (a) High-angle annular dark-field scanning transmission electron microscope (HAADF-STEM) image of an 8CdS/CdSe/8CdS NPL, viewed from the side. Scale bar is 3 nm. (b) Line profile from the HAADF-STEM image: the sharp lines correspond to the positions of Cd planes; the gaps between the sharp lines correspond to S planes in the shell region and Se planes in the core region. (c) Elemental mapping by energy dispersive x-ray (EDX) spectroscopy along the red line in (a), showing Se in the core region, S in the shell region, and Cd in both regions.

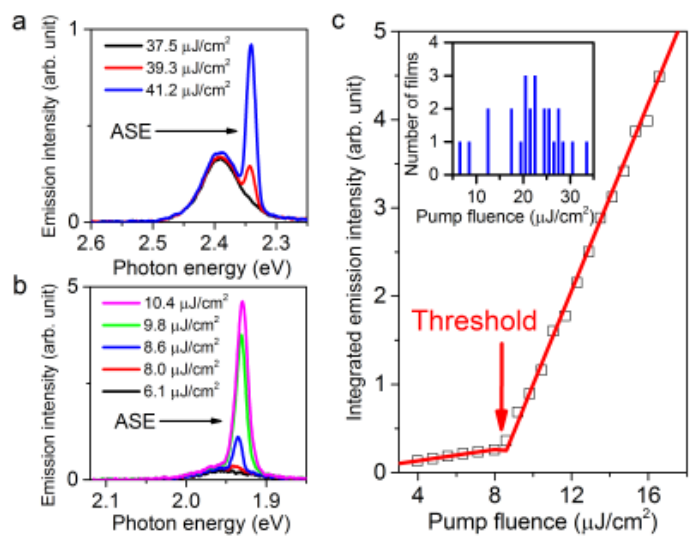

**Figure 3. Amplified spontaneous emission (ASE) from NPL films.** (a-b) Emission spectra from films of (a) CdSe and (b) 3CdS/CdSe/3CdS NPLs for different pump fluences. (c) Normalized integrated emission intensity from the film of 3CdS/CdSe/3CdS NPLs as a function of pump fluence. The dots are experimental data, and the red lines are two linear fits for different regions of pump fluence. The inset shows a histogram of the ASE thresholds for different films of 3CdS/CdSe/3CdS NPLs; each film showed similar thresholds from multiple spots.

Even lower thresholds are obtained for the NPL heterostructures. Figure 3b-c shows representative data for a film of 3CdS/CdSe/3CdS NPLs with a thickness of 135 ± 13 nm, a refractive index at 650 nm of 1.9 ± 0.1, and thus a packing density of 64±6%. For this film, the ASE threshold is 8.6 μJ/cm$^2$. Similar results were obtained for NPLs with different thicknesses of CdS shell, with no clear dependence of the threshold on shell thickness (see Supplementary Figure S3); we therefore focus in the following on 3CdS/CdSe/3CdS NPLs. The inset of Figure 3c summarizes the results from 25 films. Out of these, 2 show thresholds below 10 μJ/cm$^2$, and the best film shows a threshold of 6 μJ/cm$^2$, more than 4 times lower than the best reported value for colloidal nanocrystals[15].

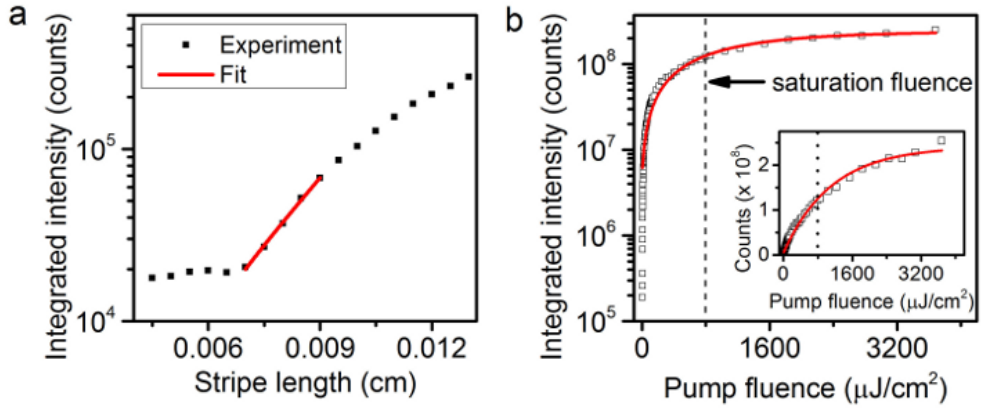

**Figure 4. Gain of NPL films.** (a) Integrated emission intensity, $I$, as a function of stripe length, $l$, from a film of 3CdS/CdSe/3CdS NPLs, for a pump fluence of 200 µJ/cm². The dots are experimental data, and the solid line is a fit to $I = I_o + A(e^{gl} - 1)/g$, where $A$ is a constant proportional to the spontaneous emission power density and $g$ is modal gain. (b) Integrated emission intensity as a function of pump fluence, showing saturation of ASE at high pump fluences. The dots are experimental data, and the red line is an exponential fit; saturation is defined as the fluence at which the intensity reaches half of its maximum value. The inset shows the same data on a linear scale.

Modal gain of the NPL films is estimated using the variable-stripe-length method[24], as illustrated in Figure 4a. A fit to the data gives a gain of 600 ± 100 cm⁻¹, approximately a factor of 4 larger than the largest values reported for QDs[7]. Moreover, the gain saturates at very high pump fluences. Figure 4b shows gain saturation at a pump fluence of 800 µJ/cm², more than two orders of magnitude higher than the ASE threshold. By comparison, gain saturation in QDs typically occurs for fluences about twice the ASE threshold[15,25].

The performance of QDs as gain media is believed to be limited by Auger recombination. We therefore used transient-absorption measurements[9] to determine Auger recombination rates in NPL solutions. As shown in Figure 5a, increasing the pump fluence results in the emergence of fast decay components, characteristic of Auger recombination[9]. By subtracting the dynamics of single excitons, we obtained Auger recombination that is non-single exponential, similar to a recent report for similar NPLs[26] in time-resolved photoluminescence measurements. The 1/$e$ decay time is 510 ± 50 ps at a fluence of 13 μJ/cm$^2$, significantly longer than the Auger lifetime for QDs with comparable emission wavelength[15]. A recent report also suggests that Auger lifeteimes for CdSe NPLs can be as long as 10 ns[26]. Slower Auger recombination may therefore be partially responsible for the reduced ASE threshold and increased gain in NPL films.

On the other hand, gain build-up times in nanocrystal films are on the order of 10 ps or less[12], shorter than even the fastest Auger recombination lifetimes; this implies that additional factors may also be responsible for the superior gain characteristics of the NPL films. One factor is likely the large absorption cross-sections of the NPLs at the pump-photon energy (3.1 eV): we measure a cross-section of 3.1×10$^{-14}$ cm$^2$ for CdSe NPLs. This is further increased by growing the CdS shell (see Supplementary Figure S4): 3CdS/CdSe/3CdS NPLs have an absorption cross-section of 1.1×10$^{-13}$ cm$^2$, which helps explain their lower ASE threshold compared to the CdSe NPLs. The shell further reduces the threshold by increasing the luminescence quantum yield of the NPLs[27].

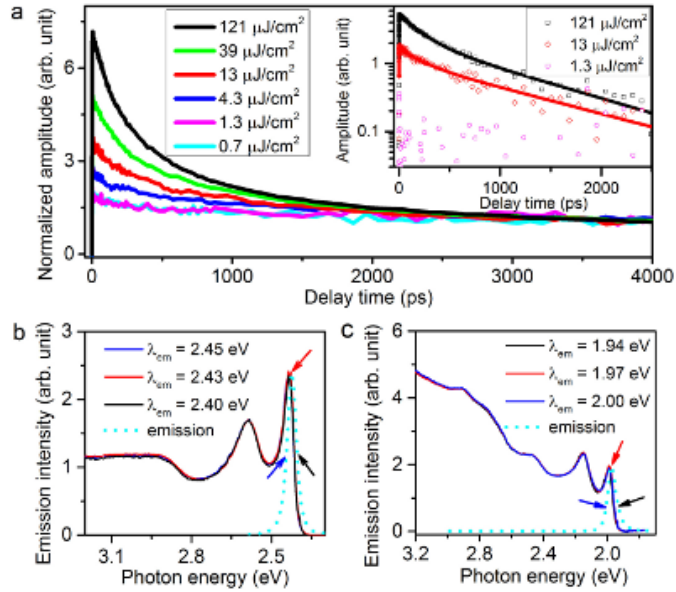

**Figure 5. Characteristics of NPLs responsible for low thresholds and high gain.** (a) Dynamics of the lowest-energy excitons for 3CdS/CdSe/3CdS NPLs in hexane, for different pump fluences. The samples are excited at 3.1 eV, and their transient absorption is probed at 1.98 eV; the signals are normalized to their values at long delay times. Inset shows examples of the dynamics obtained by subtracting the normalized data for a pump fluence of 0.7 μJ/cm$^2$. These multi-exponential short-time dynamics correspond to Auger recombination, with 1/$e$ decay times of 420 ± 30 ps and 510 ± 50 ps at 121 μJ/cm$^2$ and 13 μJ/cm$^2$, respectively. (b-c) Emission spectrum (dashed line) and photoluminescence excitation (PLE) spectra (solid lines) for (b) CdSe and (c) 3CdS/CdSe/3CdS NPLs in hexane. For each sample, PLE spectra were recorded for three different emission energies, indicated by the three arrows; the identical spectra indicate the absence of inhomogeneous broadening.

Another important factor is the absence of inhomogeneous spectral broadening in the NPLs. The synthesis of NPLs and the subsequent c-ALD growth of the shells result in an ensemble of atomically flat platelets that all have the same thickness, within a single ML, and thus all have the same emission energy. This can be seen in the identical photoluminescence excitation (PLE) spectra of the NPL solutions for different emission energies[23], shown in Figure 5b-c. If the NPL ensemble were inhomogeneously broadened, monitoring emission at a particular wavelength would select a particular sub-ensemble, and the PLE spectra would vary depending on emission energy. The absence of inhomogeneous broadening means a narrow ensemble emission linewidth, which in turn means a large gain coefficient per exciton at the ASE wavelength.

Finally, the higher gain saturation can be understood in terms of the higher band-edge exciton density that can be obtained in NPLs as compared to QDs. Once the two band-edge states in a QD are occupied, additional excitation leads to the occupation of excitons in higher-lying quantum-confined states[14]. By contrast, a continuous density of states is available at the band edge for the one-dimensionally confined excitons in NPLs, allowing for much higher occupation numbers[26].

In conclusion, colloidal nanoplatelets are in all respects a superior medium for light amplification and optical gain than conventional colloidal nanocrystals. Films of NPLs show lower ASE thresholds and high saturated gain, which can ultimately be traced to the unique synthetic process used to make the NPLs and the one-dimensional confinement of carriers in the NPLs. Our NPL films already outperform the best reported films of colloidal QDs, and a further reduction in ASE threshold is likely possible by reducing non-uniformities and scattering in the NPL films[14,28]. The use of NPLs thus has the potential to turn colloidal-nanocrystal lasers into a practical reality, combining the advantages of quantum wells with the advantages of a colloidal system. Key among these latter advantages is solution processability, which opens up the possibility of integrating low-cost lasers into nearly any system, including flexible substrates, optical fibers, microfabricated waveguides, and lab-on-a-chip systems.

**Methods**

*Sample preparation*. All sample synthesis and preparation was performed inside an $N_2$-filled glovebox.

CdSe NPLs were synthesized following a previously published procedure[29] with slight modifications. Specifically, in a three-necked flask, 170 mg of cadmium myristate was degassed in 15 mL of 1-octadecene (ODE) at 90°C for 30 min before adding 12 mg of selenium powder. The resulting mixture (cadmium myristate and Se in ODE) was further degassed at 90°C for 1 hour. We then heated the mixture rapidly and added 40 mg of cadmium acetate dehydrate at 195°C. After keeping the reaction mixture at 240°C for 5 min, we rapidly cooled the mixture down to 70°C and then injected a solution of 2 mL of oleic acid in 10 mL of anhydrous hexane. The mixture was then centrifuged, and the precipitate containing the nanoplatelets was suspended in hexane.

CdS/CdSe/CdS shell/core/shell NPLs were synthesized following one of the variants of the colloidal atomic layer deposition approach[23], with some modifications and optimizations. Before shell growth, we precipitated CdSe NPLs, synthesized as described above, and re-dispersed them in hexane three times to remove free $Cd^{2+}$ in the solution. Then, the first layer of $S^{2-}$ was introduced by phase transferring 4CdSe NPLs from hexane to 5 mL of N-methylformamide (NMF), in which 50 μL of aqueous solution of ammonium sulfide (40%) was dissolved. After phase separation, we added acetonitrile and toluene to precipitate the NPLs, and dispersed them in 5 mL of NMF. To grow the first layer of Cd, we redispersed the solution in 2 mL of NMF, introduced 2.5 mL of 0.2 M cadmium acetate in NMF, and stirred the solution for 1 minute. Then, we precipitated the NPLs with toluene, centrifuged them, and redispersed the precipitate in 5 mL of NMF. The first monolayer of CdS shell growth is completed at this stage, with the surfaces terminated by Cd layers. To grow a thickness of $x$ monolayers of CdS, the above steps were repeated $x$ times. The final shell/core/shell NPLs were dispersed in 5 mL of hexane with the addition of 250 μL of dried 70% technical-grade oleylamine.

In order to deposit the NPL films, we first precipitated the NPLs from the above solution using ethanol to remove excess oleylamine. We then redispersed the NPLs in 1 mL of a 4:1 (v/v) hexane:octane mixture and filtered the solution through a membrane filter. The resulted solution was then spin coated on a glass substrate.

*Electron-microscope imaging.* The core/shell structure was investigated using a 200 kV aberration corrected STEM (JEOL 2200FS) with a high-angle annular dark field (HAADF) detector. The intensity of the HAADF-STEM image is related to the element (Z number), density, and thickness of the imaged material[30]. In order to avoid contamination during the acquisition at high resolution, the "beam shower" procedure was performed by scanning the sample with a defocused beam at a magnification of 50 kX. Energy dispersive x-ray spectroscopy (EDX) was performed using the STEM nano-probe mode, in order to obtain a compositional line profile.

*Measurement of film properties*. The thickness and refractive index ($n_{sample}$) of the NPL films were measured using an ellipsometer (Gaertner Scientific LSE-WS, wavelength of 632 nm for CdSe NPL films; Horiba Jobin Yvon UVISEL, wavelength of 650 nm for CdS/CdSe/CdS NPL films). Average values were determined over areas 10 mm in diameter. From these values, the packing density was calculated according to $D = (n_{sample}-n_{air})/(n_{CdSe}-n_{air})$, with $n_{air}$ taken to be 1 and $n_{CdSe}$ taken to be 2.4.

*Measurement of ASE.* The films were excited with pulses obtained by frequency doubling the output of an amplified Ti:Sapphire laser system (Spectra-Physics Spitfire Pro). The pulses had a photon energy of 3.1 eV, a duration of approximately 35 fs, and a repetition rate of 100 Hz. The laser beam was focused onto a stripe along the NPL films using a cylindrical lens. The width of the stripe, defined as the width that contains 60% of the power in the laser beam, was determined using the knife-edge technique. In a typical ASE experiment, the stripe was 1.8 mm in length and 70 μm in width. In order to ensure the accuracy of our estimates of power density, we varied the stripe width in the range from 55 μm to 250 μm, and found that the measured ASE threshold was unchanged for the same sample. Emission was measured from the edge of the films, in the direction of the strip and perpendicular to the propagation direction of the incident pump beam. The collected emission was sent through a grating spectrometer and detected using a charge-coupled device (CCD) detector. All measurements are made at room temperature in air.

*Transient absorption spectroscopy.* Ultrafast transient absorption measurements were carried out using a commercial system (Ultrafast Systems HELIOS). An amplified Ti:Sapphire pulse (800 nm, 35 fs, 1 kHz repetition rate Spectra-Physics Spitfire Pro) was split into two beams. The first beam, containing 10% of the power, was focused into a sapphire window to generate a white light continuum (440 nm – 750 nm), which serves as the probe. The other beam, containing 90% of the power, was sent into an optical parametric amplifier (Spectra-Physics TOPAS) to generate the pump beam. After the pump beam passes through a depolarizer, it was focused and overlapped with the probe beam at the sample.

*Determination of absorption cross-section*. We determined the absorption cross-section of NPLs by measuring their concentration and absorbance in solution. The NPL concentration was determined by dissolving the platelets in a nitric-acid solution and measuring the concentration of Se in the solution, [Se], using inductively coupled plasma optical emission spectroscopy (ICP-OES). The dimensions of the CdSe NPLs or NPL cores were measured from TEM and HAADF-STEM images. We then calculated the concentration of NPLs according to [NPL] = [Se] $V_{unit}/(4V_{NPL})$, where $V_{unit}$ is the volume of the CdSe unit cell, and $V_{NPL}$ is the physical volume of CdSe in the core. The absorption cross-section is then determined by measuring the absorbance, $A$, of the solution, and using the following equation: $\sigma_{abs} = (2303A)/([NPL]N_A L)$, where $N_A$ is Avogadro's number and $L$ is the optical path length used in the absorbance measurement.

**Acknowledgments**

This work was performed, in part, at the Center for Nanoscale Materials, a U.S. Department of Energy, Office of Science, Office of Basic Energy Sciences User Facility under Contract No. DE-AC02-06CH11357. This work was supported by the Keck Foundation and by the University of Chicago and the Department of Energy under Department of Energy Contract No. DE-AC02-06CH11357, awarded to UChicago Argonne, LLC, operator of Argonne National Laboratory. D.V.T. thanks the David and Lucile Packard Fellowship. This work used facilities supported by University of Chicago NSF MRSEC Program under Award Number DMR-0213745. A portion of this research was conducted at the Center for Nanophase Materials Sciences, which is sponsored at Oak Ridge National Laboratory by the Scientific User Facilities Division, Office of Basic Energy Sciences, U.S. Department of Energy.

**Author Contributions**

M.P. conceived the project. D.T. and M.P. coordinated the project. I.F. and D.S.D. synthesized the nanoplatelet samples and deposited the films. C.S. and I.F. characterized the nanoplatelet samples. A.D. performed HAADF-STEM imaging and EDX spectroscopy. C.S. and R.D.S. performed optical measurements. C.S., R.D.S., I.F., M.P., and D.V.T analyzed the data. C.S., M.P., and D.V.T. wrote the manuscript, with contributions from all the authors.

**Competing Financial Interests**

The authors report no competing financial interests.

# Supplementary Information for

# Low-threshold stimulated emission using colloidal quantum wells

Chunxing She, Igor Fedin, Dmitriy S. Dolzhnikov, Arnaud Demortière, Richard D. Schaller, Matthew Pelton, and Dmitri V. Talapin

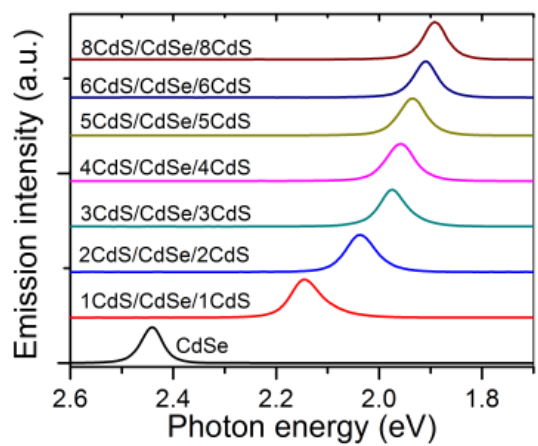

**Supplementary Figure S1**. Emission spectra of CdSe and $x$CdS/CdSe/$x$CdS nanoplatelets (NPLs) in hexane.

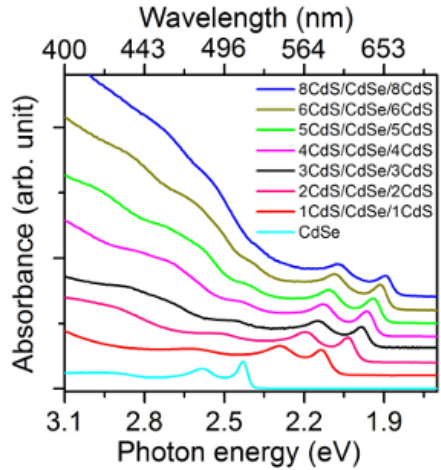

**Supplementary Figure S2**. Absprotion spectra of CdSe and $x$CdS/CdSe/$x$CdS nanoplatelets (NPLs) in hexane.

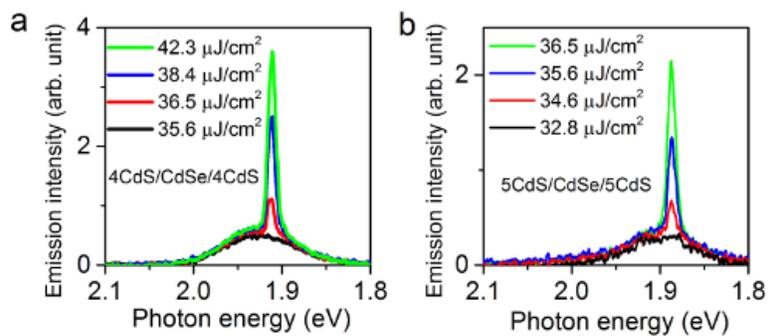

**Supplementary Figure S3.** Emission spectra from films of (a) 4CdS/CdSe/4CdS and (b) 5CdS/CdSe/5CdS NPLs for excitaton with ultrafast laser pulses focused to a stripe, showing amplified spontaneous emission.

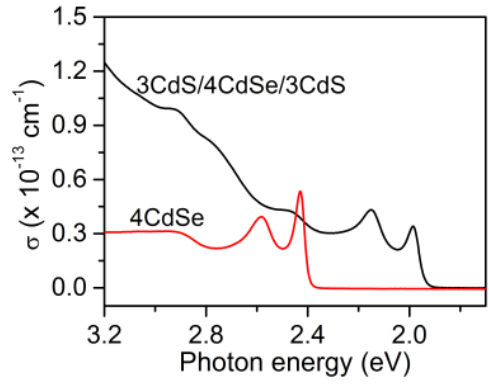

**Supplementary Figure S4**. Absorption cross-section spectra of CdSe and 3CdS/CdSe/3CdS NPLs in solution.